\documentclass[useAMS,usenatbib]{mn2e}
\usepackage{graphicx}
\usepackage{amssymb}
\usepackage{lscape}
\usepackage{ulem}
\usepackage{txfonts}

\def\kms{km~s$^{-1}$}

\def\si2{Si\,{\sc ii}}
\def\mg2{Mg\,{\sc ii}}
			\def\fe2{Fe\,{\sc ii}}
\def\al2{Al\,{\sc ii}}
\def\zn2{Zn\,{\sc ii}}
\def\c2s{C\,{\sc ii}$^{\star}$}
\def\hkpc{$h_{70}^{-1}$ kpc}

\title[Galaxy Pairs in the Sloan Digital Sky Survey II]
{Galaxy Pairs in the Sloan Digital Sky Survey - II: The Effect of 
Environment on Interactions.}

\author[Ellison et al.] {Sara L. Ellison$^1$, David R. Patton$^2$,  
Luc Simard$^3$,  Alan W. McConnachie$^{3}$, 
\newauthor Ivan K. Baldry$^4$, J. Trevor Mendel$^1$\\
$^1$ Department of Physics and Astronomy, University of Victoria, Victoria, British Columbia, V8P 1A1, Canada.\\
$^2$ Department of Physics \& Astronomy, Trent University, 
1600 West Bank Drive, Peterborough, Ontario, K9J 7B8, Canada.\\
$^3$ National Research Council of Canada,
Herzberg Institute of Astrophysics, 5071 West
Saanich Road, Victoria, British Columbia, V9E 2E7, Canada\\
$^4$ Astrophysics Research Institute, Liverpool John Moores University,
Twelve Quays House, Egerton Wharf, Birkenhead, CH41 1LD, UK. 
}

\begin{document}

\maketitle

\begin{abstract}
We use a sample of close galaxy pairs selected from the Sloan Digital
Sky Survey Data Release 4 (SDSS DR4) to investigate in what
environments galaxy mergers occur and how the results of these mergers
depend on differences in local galaxy density.  The galaxies are
quantified morphologically using two-dimensional bulge-plus-disk
decompositions and compared to a control sample matched in stellar
mass, redshift and local projected density.  Lower density
environments have fractionally more galaxy pairs with small projected
separations ($r_p$) and relative velocities ($\Delta v$), but even
high density environments contain significant populations of pairs
with parameters that should be conducive to interactions.  The
connection between environment and $\Delta v$ also implies that the
velocity selection of a pairs sample affects (biases) the environment
from which the pairs are selected.  Metrics of asymmetry and colour
are used to identify merger activity and triggered star formation.
The location of star formation is inferred by distinguishing bulge and
disk colours and calculating bulge fractions from the SDSS images.
Galaxies in the lowest density environments show the largest changes
in star formation rate, asymmetry and bulge-total fractions 
at small separations, accompanied by  bluer
bulge colours.  At the highest local densities,
the only galaxy property to show an enhancement in the closest pairs
is asymmetry.  We interpret these results as evidence that whilst
interactions (leading to tidal distortions) occur at all densities,
triggered star formation is seen only in low-to-intermediate density
environments.  We suggest that this is likely due to the typically
higher gas fractions of galaxies in low density environments.
Finally, by cross-correlating our sample of galaxy pairs with a
cluster catalogue, we investigate the dependence of interactions on
clustercentric distance.  It is found that for close pairs the
fraction of asymmetric galaxies is highest in the cluster centres.

\end{abstract}

\section{Introduction}

Galaxies in high density environments have been shown to be
unequivocally different to their isolated counterparts.  From the
discovery of the morphology-density relation (e.g. Dressler 1980;
Postman \& Geller 1984), the colours, morphologies and mean star
formation rates of galaxies all clearly depend on their local
environment (e.g Kauffmann et al. 2004; Balogh et al. 2004; Baldry et
al. 2006; Weinmann et al. 2006; Park et al 2007; 
Skibba et al.  2009; Bamford et al. 2009 and references therein).
Disentangling the process(es) that drive these environmental
dependences, the scales on which environment matters and identifying
which are the primary changes in galaxy properties is a major theme in
contemporary astrophysics (e.g. Blanton \& Berlind 2007; Park \& Choi
2009; Deng et al. 2009).

In this paper, we tackle the specific issue of tidal interactions
between galaxies and how gravitationally induced effects, such as
triggered star formation and morphological transformation, depend on
the larger scale environment.  It has been well known for many years
that galaxy-galaxy interactions can trigger star formation, e.g.
Kennicutt et al., (1987); Barton, Geller \& Kenyon (2000); Lambas et
al (2003); Alonso et al. (2004); Nikolic, Cullen \& Alexander (2004).
The enhanced star formation that accompanies
galaxy-galaxy interactions appears to be relatively modest, around a
factor of two on average, and many close pairs have `normal' star
formation rates (e.g. Barton et al. 2000; Bergvall et al. 2003; Lin et
al. 2007; Darg et al.  2010; Knapen \& James 2009; Robaina et al.
2009).  However, the efficiency with which star formation is triggered
depends strongly on not only internal, but also the relative, macroscopic
properties of galaxies in interactions.  Major galaxy-galaxy
interactions (where the ratio of galaxy masses or luminosities is
close to unity) trigger the strongest star formation (Woods, Geller \&
Barton 2006; Ellison et al. 2008).  In minor (unequal mass) pairs, the
lower mass galaxy appears to be more affected by triggered star
formation than the more massive counterpart (Donzelli \& Pastoriza
1997; Woods \& Geller 2007; Ellison et al. 2008; Li et al. 2008), as
predicted by simulations (Bekki, Shioya \& Whiting 2006; Cox et
al. 2008).  Simulations also predict that the orientation and relative
rotation directions can dramatically affect the efficiency of a
starburst, much more so than simply the presence of an adequate gas
supply (Di Matteo et al. 2007; Cox et al. 2008).  The efficiency may
also increase at higher redshifts (e.g. Bridge et al. 2007) where
some of the most luminous galaxies in the UV are seen to have close
companions and exhibit signs of interactions (Basu-Zych et al. 2009).

The triggered star formation is expected to occur in the nuclear
regions of the galaxy as gas loses angular momentum during the
interaction, modulated by the presence of a bulge (Mihos \& Hernquist
1994, 1996; Cox et al. 2008).  This prediction is supported
observationally by lower nuclear metallicities (Kewley et al. 2006;
Ellison et al. 2008), blue central colours and larger H$\alpha$
equivalent widths (Barton et al. 2003; Bergvall et al. 2003; Kannappan
et al. 2004) and higher concentration indices and bulge fractions
(Nikolic et al. 2004; Li et al. 2008; Perez et al. 2009b) in close
pairs.  Recently, Soto \& Martin (2009) have added further evidence to
this scenario from the detection of age gradients in Ultra Luminous
Infra-Red Galaxies (ULIRGs) that are consistent with gas removal from
the outer regions to fuel continuous star formation in the centres.
In addition to a population of very blue galaxies, and examples of highly
efficient star formation, close pairs also appear to have a notable
red population (Alonso et al. 2006; Darg et al. 2010; Perez et al.
2009b).  Triggered star
formation in mergers, although prevalent in late-type galaxies, is not
significant in early-type pairs, probably due to their low gas
fractions (Luo, Shu \& Huang 2007; Park \& Choi 2009; Darg et
al. 2010; Rogers et al. 2009).

A number of previous works have assessed the role of local environment
on galaxy-galaxy interactions.  Lambas et al. (2003) and Alonso et al.
(2004) studied 2-degree field (2dF) pairs in the field and in groups
respectively.  It was found that in groups, enhanced star formation
required smaller separations than in the field.  Alonso et al. (2006)
extended the 2dF analysis to include pairs in the Sloan Digital Sky
Survey (SDSS) and compared the birth rate parameter ($b$, proportional
to the specific star formation rate) in low, intermediate and high
density environments, using the 5th nearest neighbour ($\Sigma_5$) as
a density estimator for galaxies in the SDSS.  Enhanced star formation
was seen in wider separation pairs for lower $\Sigma_5$.  Although
enhanced star formation and bluer colours were found most
significantly for pairs in the lowest density environments, the
results may be biased by the fact that Alonso et al. (2006) did not
match their control sample equally in $\Sigma_5$ (Perez et al. 2009a).
Perez et al. (2009b) have recently improved upon this analysis, by
constructing samples of pairs and control galaxies matched in stellar
and dark matter halo mass,
redshift and $\Sigma_5$.  It was found that although pairs in the
lowest density environments have the highest absolute fraction of star
forming galaxies, pairs with projected separations $r_p < 20$ kpc have
approximately twice the fraction of high $b$ galaxies compared to the
control, regardless of local density. However, Perez et al.  (2009b)
report that the colour and concentration distributions of the pairs
and control differ most at intermediate densities (but see Patton et al
in prep and Simard et al. in prep. for caveats on the use of SDSS
photometry in crowded environments).  Therefore,
although the enhanced star formation in low density environments is
likely the cause of a tail of very blue galaxies, the effect on the
overall colour distribution is minor.  Perez et al.  (2009b) suggested
that intermediate density environments, perhaps synonymous with
groups, are the most important environment for galaxy interactions and
mergers.

We are investigating the effects of environment on galaxy evolution by
looking at galaxies in a range of environments; clusters (Ellison et
al. 2009), compact groups (McConnachie et al. 2008, 2009; Brasseur et
al. 2008) and pairs, of which this is the second paper in the series.  In
Paper I (Ellison et al. 2008) we focussed on a sample with strong
emission lines to investigate the effect of galaxy interactions on the
mass-metallicity relation and fraction of active galactic nuclei.  In
two forthcoming papers, Patton et al. (in preparation) use a larger
sample to study the global colours of galaxy pairs, and Simard et
al. (in preparation) examine the detailed morphological properties of
pairs.  In this paper we re-visit the issue of mergers and triggered
star formation as a function of environment by considering the effect
of both projected density and cluster membership on the same sample of
close pairs.  This is important in order to disentangle different
physical processes and dependences.  For example, the star formation
rates (e.g. Gomez et al. 2003; Lewis et al. 2002) and mass-metallicity
relation (Ellison et al. 2009) of galaxies are apparently dictated
more by local density than cluster membership, possibly because these
effects are driven by galaxy-galaxy interactions.  On the other hand,
the fraction of post-starburst galaxies is more dependent on cluster
or group membership (e.g. Poggianti et al.  2009), indicating that ram
pressure stripping may be important (Ma \& Ebeling 2009).

The layout of the paper is as follows.  In Section \ref{sample_sec} we
describe the selection of our pairs sample and the construction of a
matched control sample.  A properly designed control sample is
essential for disentangling the effects of galaxy-galaxy interactions
from biases in the pairs selection, such as projection effects (
e.g. Perez et al. 2009a) or photometric errors.  In Section
\ref{sigma_sec} the dependence of pairwise properties as a function of
environment, as defined by a 2-dimensional projected density, are
investigated.  The colours, asymmetry and bulge fractions of the close
pairs sample are then analysed as a function of projected separation
and environment.  The importance of cluster membership is investigated
in Section \ref{member_sec}.

We adopt a concordance cosmology of $\Omega_{\Lambda} = 0.7$, $\Omega_M = 0.3$
and $H_0 = 70$ km/s/Mpc.

\section{Sample selection and characterization}\label{sample_sec}

The galaxy sample used in this work is selected from the SDSS Data
Release (DR) 4 and uses similar criteria to those defined by Patton et
al. (in prep.).  We briefly review the salient selection criteria
here, highlighting the distinction with the other papers in this
series (Ellison et al. 2008; Patton et al. in prep; Simard et al. in
prep.).

\begin{enumerate}

\item Extinction corrected Petrosian magnitudes must be in the range
14.5 $< r$ $\le$ 17.77.  Objects must be classified as galaxies from
SDSS imaging and in the SDSS spectroscopic catalogue
(\textit{SpecPhoto.SpecClass=2} and \textit{SpecPhoto.Type=3}).

\item Galaxies must be unique spectroscopic objects and the SDSS
SpecObjAll parameter which measures the redshift confidence must have
a value exceeding $zconf > 0.7$.

\item In contrast to the sample of Patton et al. (in prep.), we impose an
upper redshift cut-off of $z=0.1$, because cluster membership
(in this work, we use the catalogue of von der Linden et al. 2007) and
local densities have only been determined up to this distance.

\item Due to incompleteness in the SDSS spectroscopic sample incurred
by a minimum fibre separation of 55 arcsecs, wide separation pairs are
preferentially selected at $z < 0.08$ (see Figure 1 of Ellison et
al. 2008). Ellison et al. (2008) therefore removed a fraction of the
wide separation pairs to yield a sample
unbiased in the redshift-projected separation plane.  Without this
culling procedure, the distribution of mass ratios is slightly more
peaked around equal mass pairs, due to the preferential omission of
pairs with highly disparate masses.  This effect is well-known and
attributed to the loss of dynamical range in magnitude limited samples
(e.g., Patton \& Atfield 2008).  The same culling procedure is
used here.

\item The morphology of the galaxy must be successfully modelled as a
bulge plus disk in the $g$- and $r$-band with the Galaxy Image 2D
(\textsc{GIM2D}) software (see Section \ref{gim2d_sec}).  
The percentage of fits that
failed is small -- 0.17\%.  The details of the \textsc{GIM2D} software
are documented in Simard et al. (2002).  Unless otherwise stated, we
use the rest frame magnitudes derived from \textsc{GIM2D} with
k-corrections applied from Blanton \& Roweis (2007).

\item Relative stellar masses of the two galaxies in a pair (taken
from Kauffmann et al. 2003) must be within a factor of 10.  See
Ellison et al. (2008) for a discussion of selection based on relative
flux rather than mass.

\item The projected separation between a galaxy and its companion must
be within $r_p < 80$ \hkpc.

\item The rest-frame velocity difference of a galaxy pair must be
$\Delta v < 500$ \kms.

\end{enumerate}

\begin{figure}
\centerline{\rotatebox{0}{\resizebox{8cm}{!}
{\includegraphics{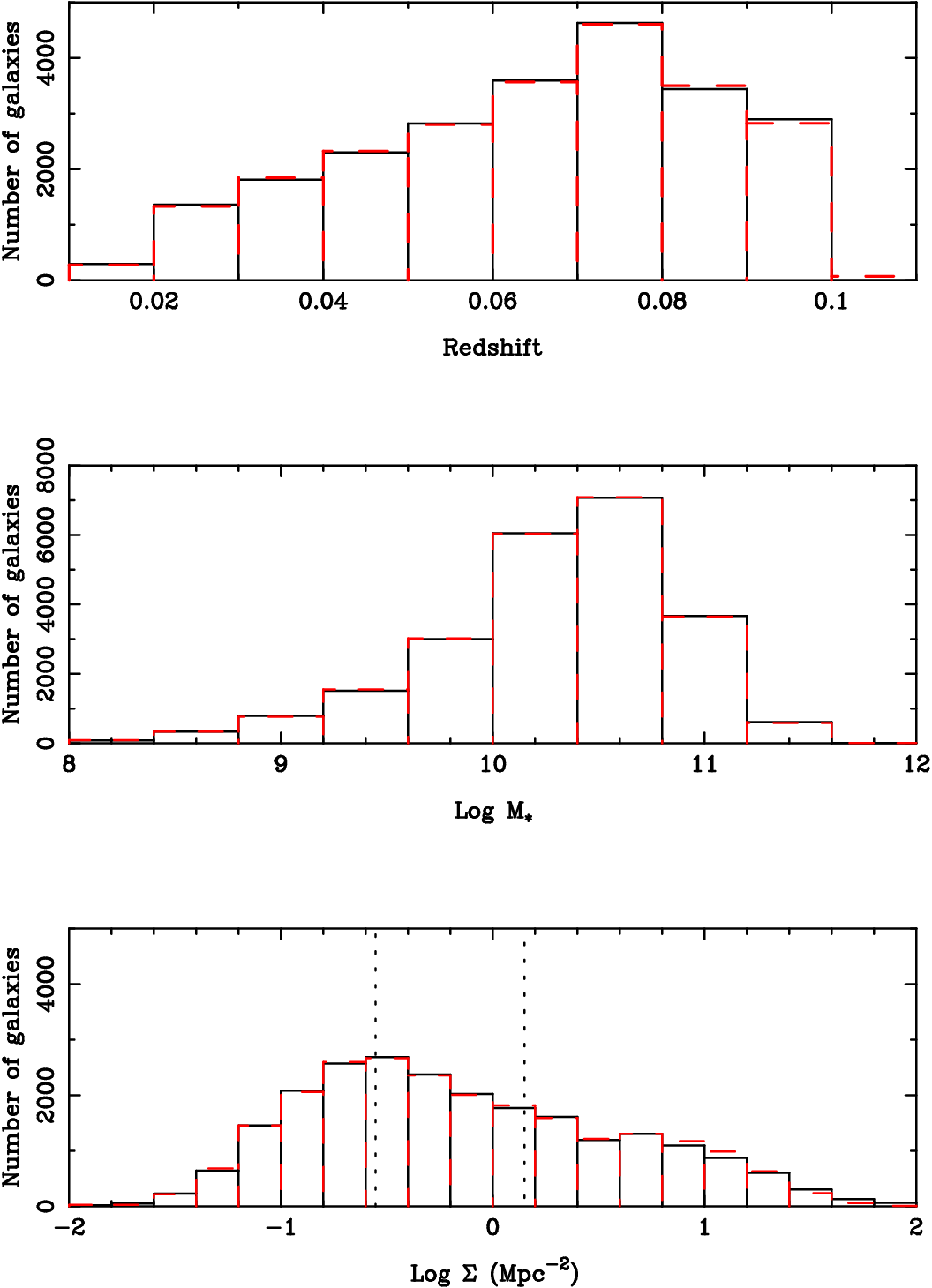}}}}
\caption{\label{sample_fig} Physical properties of the galaxy pair
sample.  Dashed red histograms show the distributions of the control
samples and solid lines represent the galaxies in close pairs.  All of
the black histograms have been scaled by a factor of four so that the
total number of galaxies shown is the same as for the control
sample. Vertical dashed lines in the lower panel indicate the
boundaries between the three tertiles in projected density.}
\end{figure}

\begin{figure*}
\centerline{\rotatebox{270}{\resizebox{13cm}{!}
{\includegraphics{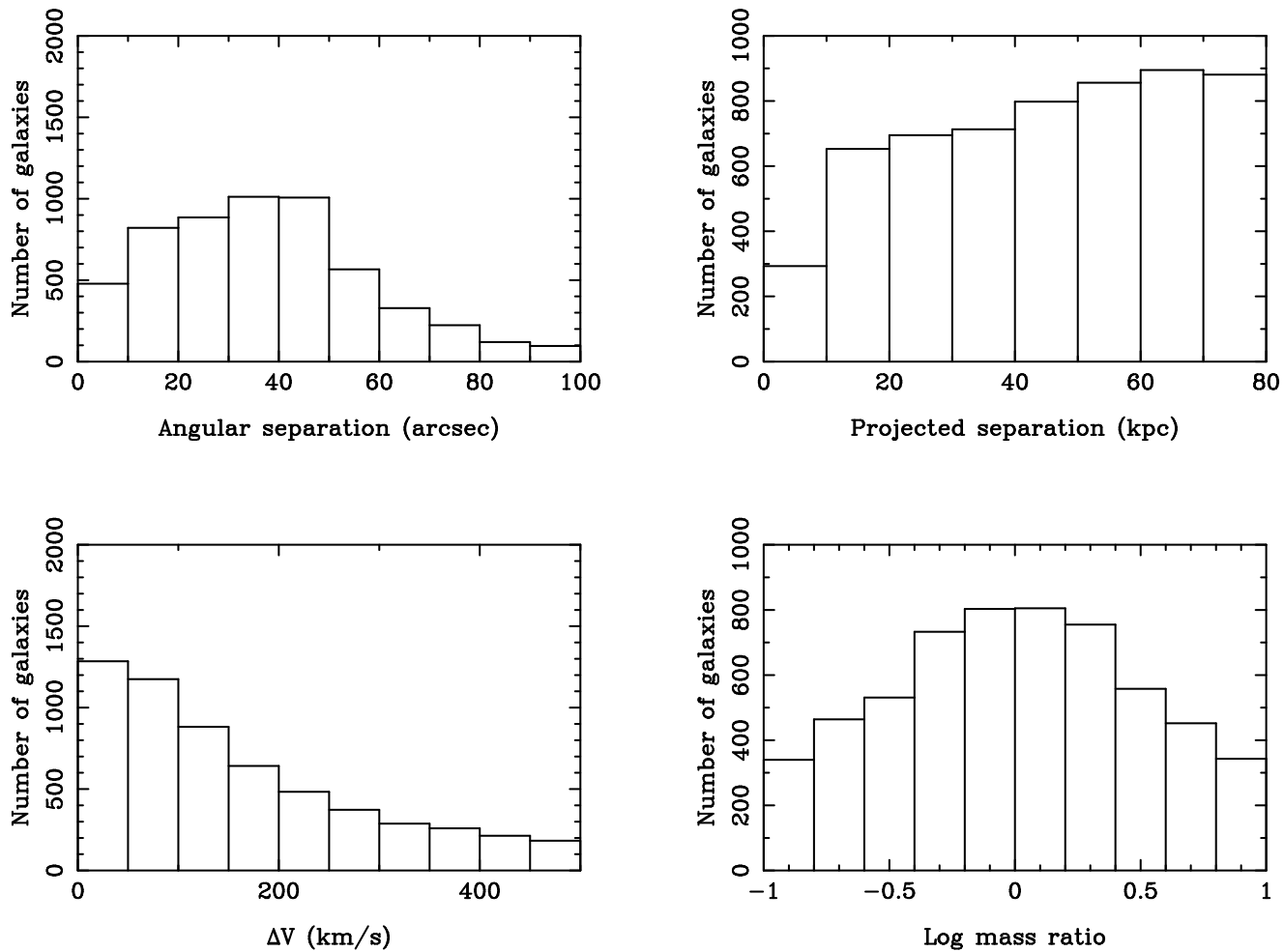}}}}
\caption{\label{pairs_fig} Pairwise properties of the overall galaxy pair
sample.   }
\end{figure*}

These criteria result in a sample of 5784 galaxies with a close
companion.  A control sample is constructed by selecting galaxies
without close companions that fulfill the first five of the above
criteria and are matched in redshift, stellar mass and environment, as
defined by

\begin{equation}\label{sigma_eqn}
\log \Sigma = \frac{1}{2} \log \left(\frac{4}{\pi d_4^2}\right) + \frac{1}{2} \log \left(\frac{5}{\pi d_5^2}\right).
\end{equation}

\noindent i.e. a projected galaxy density averaged from the distances
to the 4th and 5th nearest neighbours within 1000 \kms\ (Baldry et
al. 2006).  We adopt the densities re-computed by Ellison et
al. (2009) using a density-defining population with $M_r<-20.6$, which
yields a volume-limited sample out to $z=0.11$.  To account for
missing spectroscopic redshifts, log $\Sigma$ is also calculated using
photometric redshifts.  The final value of log $\Sigma$ is the average
of the spectroscopic redshift value and the value calculated with the
inclusion of photometric redshifts (see Baldry et al 2006 for further
discussion).  We note that the distances to the 4th and 5th nearest
neighbours are large ($\gg 500$ \hkpc) compared to the separation of
the galaxy pairs ($<$ 80 \hkpc).  
For example, the median values of
$d_4$ and $d_5$ are 1.4 and 1.8 Mpc respectively, more than an order of
magnitude larger than the separation of the pairs themselves.
The presence of a close companion therefore does not bias the use of
$\Sigma$ as a density indicator.  The control sample is constructed by
matching each pair galaxy in redshift, stellar mass and $\Sigma$ with
a galaxy with no close companion.  After matching each of the 5784
pair galaxies, the Komogorov-Smirnov (KS) probability that the control
sample is drawn from the same distribution of redshift, stellar mass
and density is calculated.  If all the probabilities are better than
30\%, the matching procedure is considered successful and is repeated.
Four control galaxies per pair galaxy were allocated before the KS
probability dropped below the 30\% tolerance level.

In Figure \ref{sample_fig} we show the distribution of pair and
control galaxy properties.  Visually, it can be seen that the control and the
pairs are well-matched in $z$, stellar mass and $\Sigma$.  The KS
probabilities that the pairs and control are drawn from the same
parent population are 99.7\% (redshift), 99.9\% (stellar mass) and 63.3\%
($\Sigma$).  Also marked on Figure \ref{sample_fig} are the values
of $\Sigma$ that correspond to equal thirds of the distribution (i.e.
tertile boundaries).  The corresponding values are $\log \Sigma = -0.55$ and
0.15.  These are close to the boundaries defined
by Perez et al. (2009b) for their low, intermediate and high density
environments (as defined by the distance to the 5th nearest neighbour).
In Figure \ref{pairs_fig} we show the pairwise galaxy properties for
our sample: projected and angular separation, $\Delta v$ and 
stellar mass ratio.

\subsection{Morphological decomposition}\label{gim2d_sec}

\textsc{GIM2D} is used to re-derive all the galaxy photometry, as well
as to determine morphological parameters.  The software is described
in detail by Simard et al. (2002).  In the work presented here, we
focus on the bulge-to-total (B/T) ratios and galaxy asymmetries.  In
brief, \textsc{GIM2D} fits a two component model (bulge plus disk) to
each galaxy image, taking into account the image point spread function
(PSF).  The ratio of flux in the bulge relative to the total galaxy
light yields the B/T ratio and can be derived separately in each
bandpass.  When multiple bands are fit, it is also possible to combine
the component solutions to yield bulge and disk colours, in addition
to integrated photometry.  Bulge and disk absolute magnitudes in
any given band are derived from

\begin{equation}
M_{\rm bulge} = M_{\rm galaxy} - 2.5 \log(B/T)
\end{equation}

\begin{equation}
M_{\rm disk} = M_{\rm galaxy} - 2.5 \log(1 - B/T)
\end{equation}

where $M_{\rm galaxy}$ is the total galaxy absolute magnitude and 
B/T is the bulge-to-total fraction (in the same band).  Asymmetry can be
quantified in many ways; we make use of the \textsc{GIM2D} parameter
$R_T + R_A$.  This metric involves subtracting the bulge plus disk
model from both the original image and a 180-degree rotated version
thereof.  $R_T + R_A$ is the sum of the residuals in these subtracted
images, normalized by the total (original) image flux.

When dealing with close pairs of galaxies, it is vital to be vigilent
for the possible effects of crowding on the photometry and
morphological decomposition.  For example, Masjedi et al. (2006) found
a dramatic increase in flux in object pairs at separations of $<$ 3
arcsecs, but extending out as far as 20 arcseconds (see also McIntosh
et al. 2008).  We have carefully studied the residual images of
\textsc{GIM2D} fits and find that the closest pairs often have poorly
fitted backgrounds and inappropriate segmentation images from SDSS
(Simard et al. in preparation).
This problem manifests itself most dramatically in the scaling
relation between disk size and luminosity where the close pairs show a
strong excess of very large disks at a given luminosity (and
consequently under-estimating the bulge fractions).  We have made
adjustments to the \textsc{GIM2D} fitting parameters to account for
crowding issues in our close pairs sample.  These include re-computing
the local object backgrounds and fitting the bulge and disk models in
the $g$ and $r$-bands simultaneously.  Full details of the fitting of
the SDSS galaxies is given in Simard et al. (in preparation), which
also presents a catalogue of the \textsc{GIM2D} parameters of $\sim$
2.2 million galaxies in the SDSS spectroscopic sample with $r < 18$.  
The improved fitting not only removes the excessive
presence of large disks, but also removes many of the extremely blue
and red objects that had been found in previous pair samples.  The
integrated colours of our pairs sample is dealt with in a separate
paper (Patton et al., in preparation).

\section{The Environment of Galaxy Pairs}\label{sigma_sec}

\begin{figure*}
\centerline{\rotatebox{270}{\resizebox{13cm}{!}
{\includegraphics{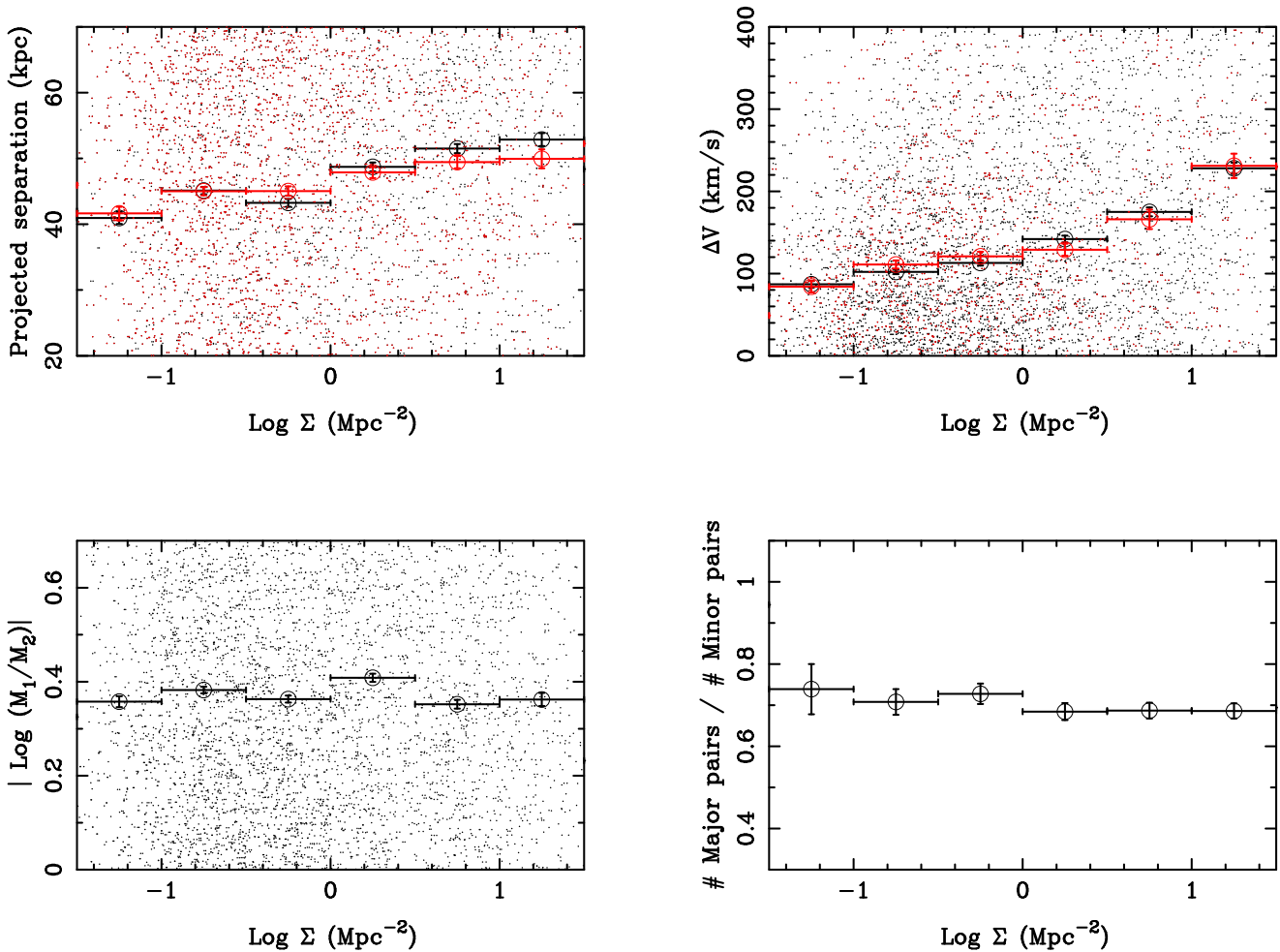}}}}
\caption{\label{pairs_sig} Pairwise properties as a function of projected
galaxy density (environment), $\Sigma$.  The individual galaxies are shown
as small dots, the binned medians and error (RMS/$\sqrt{N}$) are shown with
open circles.  Red points indicate either pair galaxies with $r_p < 30$
\hkpc\ (upper right panel) or  pair galaxies with $\Delta v < 200$
\kms\ (upper left panel). }
\end{figure*}

\begin{figure}
\centerline{\rotatebox{0}{\resizebox{8cm}{!}
{\includegraphics{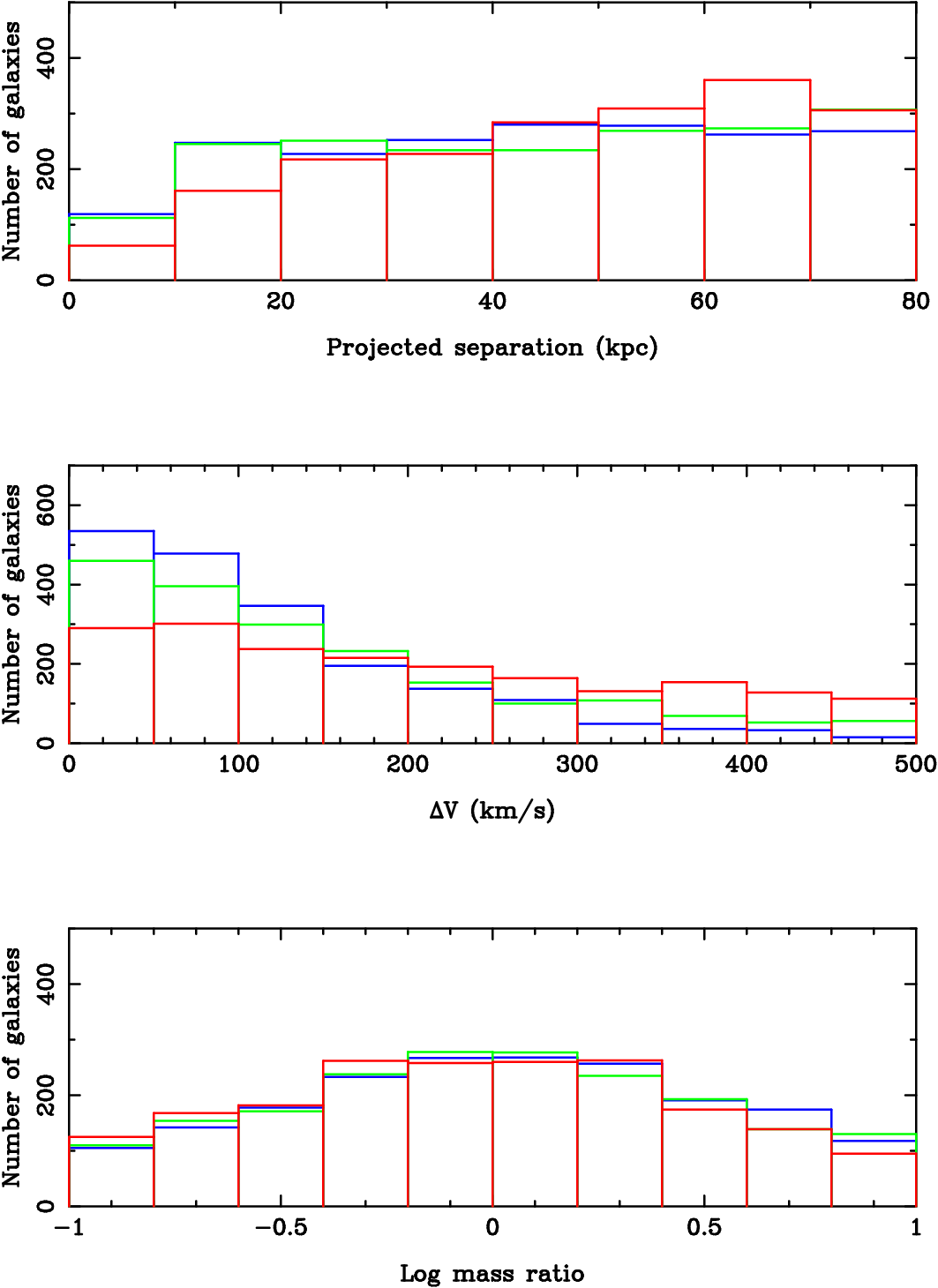}}}}
\caption{\label{pairs_sig_histo} Histograms of pairwise properties for
galaxies in high ($\log \Sigma > 0.15$, red), intermediate ($-0.55 <
\log \Sigma < 0.15$, green) and low ( $\log \Sigma < -0.55$, blue)
density environments. Although the mass ratios of pairs is independent
of environment (lower panel), lower local densities have a higher
fraction of pairs with small $\Delta v$ and smaller $r_p$.}
\end{figure}

In this section, we investigate how the local environment, as defined
by $\Sigma$, affects the selection of galaxies with a close companion,
and how such an environmental dependence influences their observed
properties.  Close galaxy pairs show increased star formation rates
when their physical separations and relative velocities are small
(e.g. Alonso et al. 2004, 2006).  Moreover, Ellison et al. (2008)
showed that star formation in galaxy pairs is more enhanced for more
equal mass pairings than for unequal mass pairings, as predicted by
simulations (e.g. Cox et al. 2008).  It is therefore germane to
investigate whether different environments host pairs whose physical
properties are more conducive to galaxy-galaxy interactions.

\subsection{Pairwise properties as a function of environment - 
where will mergers occur?}

In Figure \ref{pairs_sig} the pairwise properties of our sample are
shown as a function of $\Sigma$.  For trends of $\Sigma$ with $r_p$
and $\Delta$V, we consider both the full sample (black points) and
only those galaxies in the closest pairs ($r_p <$ 30 \hkpc, $\Delta v <200$
\kms, red points).  The mass ratio distributions are
investigated with two different metrics - the median stellar mass
ratio (the absolute value of the logarithm is used so that the primary
and secondary counterparts of a given pair are counted in the same
bin), and the ratio of major to minor pairs.  A major pair is defined
as having a ratio of stellar masses $0.5 < M_1 / M_2 < 2$.  Figure
\ref{pairs_sig} shows that the median mass ratio of pairs and the
fraction of major-to-minor pairs are independent of local environment.
However, more dense environments become increasingly dominated by
wider separation and high $\Delta v$ pairs.  Particularly noticable is
the dominant population of $\Delta v < 200$ \kms\ pairs in
environments with $-1 < \log \Sigma < 0$.  Figure
\ref{pairs_sig_histo} demonstrates the same result in a different way.
Histograms of projected separation, $\Delta v$ and mass ratio are
shown for the three $\Sigma$ tertiles (with boundaries $\log \Sigma =
-0.55, 0.15$, see Figure \ref{sample_fig}).  The mass ratio histograms
trace each other closely for the three environments, but the $\Delta
V$ and projected separation histograms are significantly different.
Low density environments ($\log \Sigma < -0.55$) are dominated by low
velocity separation pairs, with only 7\% exhibiting $\Delta v >$ 300
\kms. In contrast, 30\% of the high density ($\log \Sigma > 0.15$)
have $\Delta v >$ 300 \kms.
The projected separation distributions are also skewed, although less
dramatically.  Low density environments have a slight excess of pairs
with small separations ($r_p < 20$ \hkpc) and deficit of pairs with
wide separations ($r_p > 50$ \hkpc), relative to the pairs in high
density environments.

The corrollary of Figures \ref{pairs_sig} and \ref{pairs_sig_histo} is
two-fold.  First, low density environments may be the most common
sites for mergers, since they have a higher fraction of pairs with lower
$\Delta v$ and lower $r_p$.  34\% of the lowest density tertile have
$\Delta v < 200$ \kms\ and $r_p < 40$ \hkpc, compared to only 20\% for
the highest density tertile. Second, although a significant fraction
of galaxy pairs may be found in dense environments (Barton et
al. 2007; Perez et al. 2009b), a higher fraction of these pairs have
large relative velocities which may hinder a final merger.
Nonetheless, a significant fraction of galaxies even at high densities
have low $\Delta v$ and $r_p$, and yet galaxies in high density
environments do not seem to exhibit much triggered star formation
(Darg et al. 2010; Perez et al. 2009b).  We return to this point in
Section \ref{sfr_sec}.

It is also noteworthy that the dependence of the distribution of
$\Delta v$ on environment means that relative velocity itself is a
coarse proxy for local density.  This is not surprising, given that
more massive (high density) structures, such as clusters have larger
velocity dispersion.  This means that care must be taken when
comparing different samples of galaxy pairs, since different
tolerances of $\Delta v$ will lead to a different mix of galaxy types.
Pair samples with a lower $\Delta v$ threshold will tend to include
more later type galaxies, characterised with higher star formation
rates and smaller bulges, on average.  Moreover, relative velocities
of 1000 \kms\ are possible at peri-galacticon and Patton et al. (in
prep.)  show that close pairs exhibit evidence of interaction even
when $\Delta v \sim$ 500 -- 1000 \kms.  The trends of star formation
rate with $\Delta v$ that have previously been shown in, e.g., Alonso
et al. (2006), may therefore not be an indicator of the relative velocity
required for an interaction, but may simply trace the change in galaxy
population.  However, the frequency of truly interacting pairs is
likely to increase at smaller $\Delta v$, so that a cut-off in the
range of a few hundred \kms\ is useful for minimizing contamination.

The weak dependence of $r_p$ with $\Sigma$ is a less intuitive effect
to understand and we suggest that it may be caused by a number
of contributing factors.  At larger (projected) densities, contamination
by (non-interacting) interlopers becomes more severe.  Projection
effects preferentially contribute at larger $r_p$ and may therefore
skew the median $r_p$ to higher values at higher values of $\Sigma$
(e.g. Perez et al. 2006).
We show in the upper left panel of Figure \ref{pairs_sig} that 
this is a contributing factor by comparing the full pairs sample
($\Delta v < 500$ \kms, black points) with those whose relative
velocity is only $\Delta v < 200$ \kms\ (red points).  The lower velocity
sample will contain fewer projected interlopers, and indeed the
correlation between $r_p$ and $\Sigma$ is weaker in the $\Delta v < 200$ 
\kms\ pairs.  Another contributing factor is a redshift bias, combined
with fibre collisions.  The high $\Sigma$ sample has fewer lower
redshift galaxies and proportionately more high redshift ones.
This is due to both the limited volume sampling for rare high density
structures at low $z$ and the fact that galaxies in high density
environments tend to be more luminous and can therefore be detected
out to greater distances.  For a given $r_p$, the angular separation
on the sky is smaller at higher redshifts.  Patton \& Atfield
(2008) have shown that the spectroscopic incompleteness of the SDSS
steadily declines within the 55 arcsecond fibre collision radius,
and drops sharply (by a factor of 2) within 10 arcseconds (15 \hkpc\ 
at $z=0.08$).  This will lead to a higher incompleteness at small
physical separations for the higher density pairs.  However,
we emphasize that any dependence on
$\Sigma$ (such as redshift) is accounted for in our analysis by
matching in the control sample.

The lack of dependence of mass ratio on environment indicates that the
efficiency of triggered star formation in minor versus major pairs is
unlikely to bias observations.  

\subsection{Star formation rates}\label{sfr_sec}

\begin{figure*}
\centerline{\rotatebox{270}{\resizebox{13cm}{!}
{\includegraphics{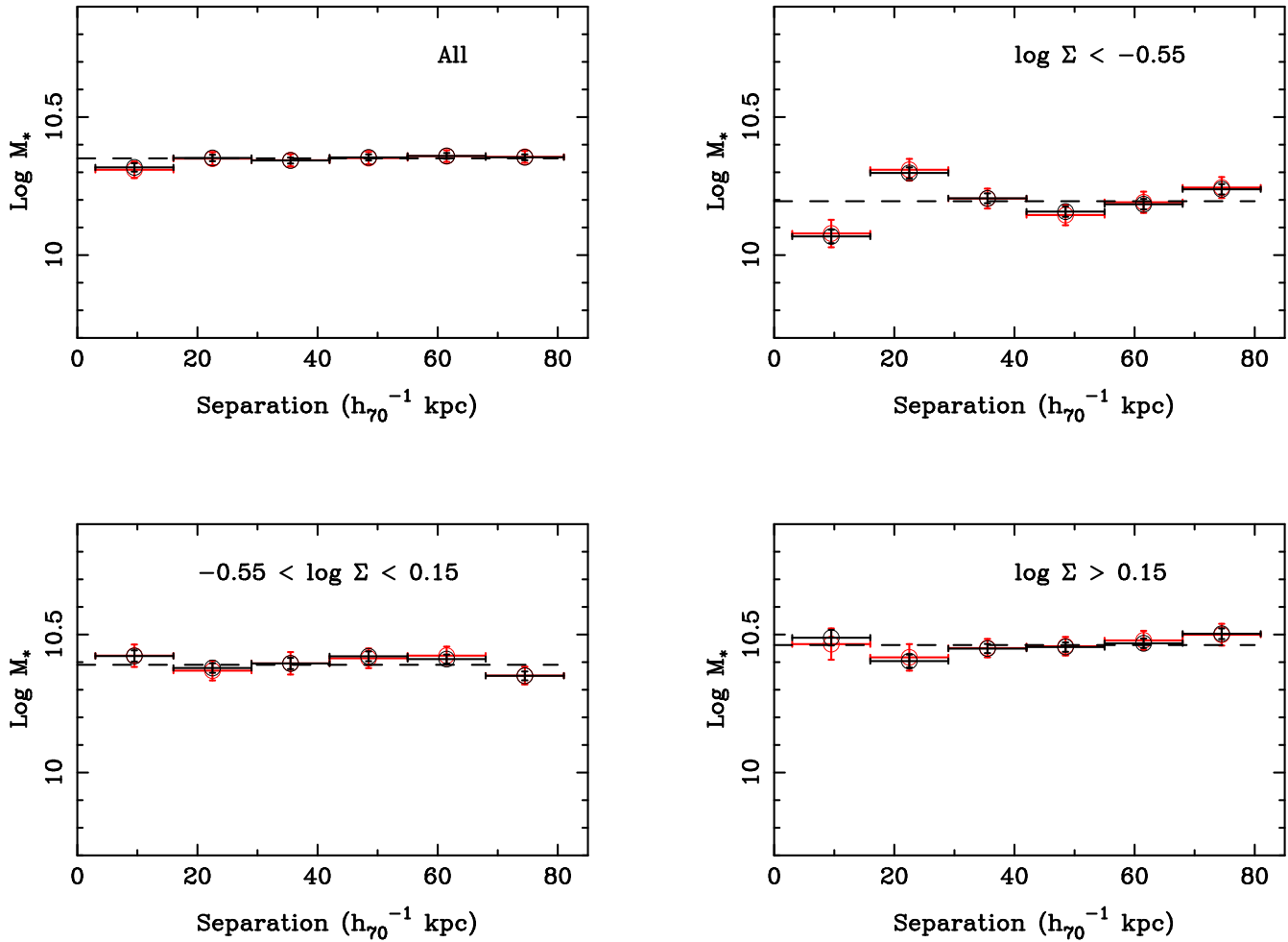}}}}
\caption{\label{mass_sep} Log M$_{\star}$ for pair galaxies
with $\Delta v < 200$ \kms\ as a function of projected separation (red
points). The
upper left panel shows the full sample, other panels are split by
local density $\log \Sigma < -0.55$ (upper right), $-0.55 < \log \Sigma
< 0.15$ (lower left) and $\log \Sigma > 0.15$ (lower right).  The
median value of stellar mass for the appropriate density control
galaxies are shown as horizontal dashed lines.  Black points show the
control galaxies in projected separation bins that correspond to the
pair to whch they are matched.}
\end{figure*}

\begin{figure*}
\centerline{\rotatebox{270}{\resizebox{13cm}{!}
{\includegraphics{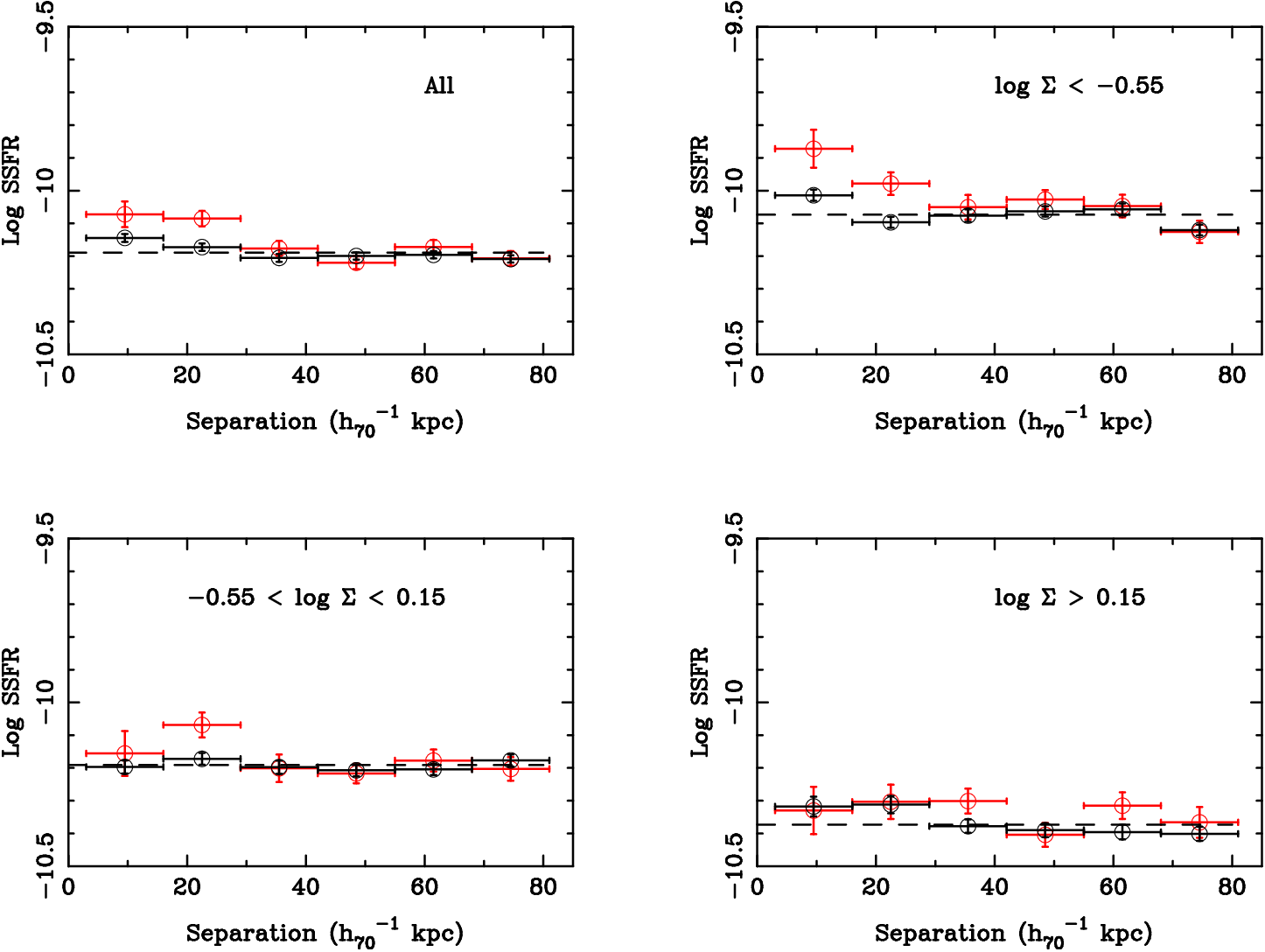}}}}
\caption{\label{sfr_sep} Specific star formation rate for pair galaxies
with $\Delta v < 200$ \kms\ as a function of projected separation (red
points). The
upper left panel shows the full sample, other panels are split by
local density $\log \Sigma < -0.55$ (upper right), $-0.55 < \log \Sigma
< 0.15$ (lower left) and $\log \Sigma > 0.15$ (lower right).  The
median value of SSFR for the appropriate density control
galaxies are shown as horizontal dashed lines.  Black points show the
control galaxies in projected separation bins that correspond to the
pair to whch they are matched.}
\end{figure*}

It is well known that enhanced star formation is seen in close pairs
of galaxies (e.g. Kennicutt et al., 1987; Barton, Geller \& Kenyon
2000; Lambas et al 2003; Alonso et al. 2004; Nikolic, Cullen \&
Alexander 2004; Woods et al. 2006; Li et al. 2008; Ellison et
al. 2008).  Before investigating how the enhancement of specific star
formation rate (SSFR) depends on environment, we first assess whether
the pairs sample may systematically vary as a function of separation,
introducing a dependence on $r_p$ that is independent of interactions.
The most important possible bias is that of stellar mass.  In Figure
\ref{mass_sep} we show the mass of galaxies in pairs (and their
matched control) as a function of separation.  The top left panel
shows that the full pairs sample is biased to slightly lower stellar
masses in the inner most separation bin, relative to the rest of the
sample.  A more striking deviation is seen in the low density sample
where pairs in the two smallest projected separation bins ($r_p < 30$
\hkpc) have median stellar masses that differ by approximately 0.1 dex
from the median of the full sample.  This is not due to small number
statistics -- there are $\sim$ 230 galaxies in each bin.
The control galaxy masses
track the pair galaxies in a given bin because stellar mass is one of
the quantities that is matched.  Figure \ref{mass_sep} highlights the
importance of control sample matching and shows that it is the offset
between the control and pairs \textit{at a given separation} that
reveals the presence of interaction-induced effects.

In Figure \ref{sfr_sep} the specific star formation rate for the pairs
sample is shown as a function of projected separation for pairs whose
$\Delta v < 200$ \kms.  The star formation rates are total,
aperture-corrected values from Brinchmann et al. (2004).  The panels
show the pairs in the full sample (upper left) and in the three
$\Sigma$ density tertiles.  In each panel, the horizontal dashed line
shows the median value for the control galaxies in the appropriate
$\Sigma$ sample.  We also show the control values binned by projected
separation, where the value of $r_p$ is taken to be that of the pair
galaxy to which it is matched.  This will allow us to see any bias
associated with pair identification as a function of separation.
Although the full pairs sample shows a 10\% increase in SSFR at small
projected separations ($r_p < 30$ \hkpc), the control galaxies matched
to the closest separation pairs also exhibits a small enhancement over
the median of the full control sample.  This is due to the slightly
lower mass of both the pairs and the control samples in the smallest
$r_p$ bin (Figure \ref{mass_sep}).  However, splitting the sample into
density tertiles reveals a much clearer enhancement in the SSFR at low
$\Sigma$ that is significant above the control out to 30 \hkpc.  At
intermediate densities, there is a significant enhancement in SSFR at
$r_p \sim$ 20 \hkpc, but not in the innermost bin.  However, the
statistics are fairly sparse at the smallest separations.  The
increase in SSFR at low and intermediate densities is consistent with
the study of Alonso et al.  (2004) who found that galaxies in both the
group and field environments could experience triggered star formation
in close pairs.  The highest density bin shows no significant trend of
enhanced SSFR relative to the control at small $r_p$.  We conclude
that only at the lowest densities is the evidence for triggered star
formation robust, with a possible enhancement at intermediate
densities.

\subsection{Bulge fractions}\label{bt_sec}

A number of works have previously found that galaxies with a close
companion have higher central concentrations (e.g. Nikolic et
al. 2004; Li et al. 2008).  Perez et al. (2009b) found that this trend
is strongest at intermediate densities, although they did not
investigate the dependence with projected separation.  Concentration
is usually defined as a ratio of the different fractional light radii
(e.g. $r_{50}$ and $r_{90}$).  With \textsc{GIM2D}, a complementary
metric can be used, namely the fraction of light that is fitted in the
bulge relative to the total flux.  An advantage of \textsc{GIM2D} over
concentration indices is that the software properly accounts for
atmospheric seeing by including the PSF in the fitting process.
Simard et al. (2002) have showed that the bulge-plus-disk
deconvolution is very robust against flux contributions from
additional components, even when 30\% of the galaxy flux is present in
superposed HII regions.  This is important when considering close
pairs where the allocation of flux between objects is crucial.
Nonetheless, it is worth highlighting the distribution of angular
separations in our sample, shown in the upper left panel of Figure
\ref{pairs_fig}.  Only 1\% of our pairs sample have angular
separations less than 5 arcseconds ($\sim$ 5 \hkpc\ at $z=0.05$).  As
discussed by Patton et al. (in prep.), galaxies with projected
separations above 5 \hkpc\ are clearly distinct in the SDSS images
(image mosaics are presented for close separation pairs in Patton et
al.).  In this and subsequent sections we show that pairs exhibit
morphological trends out to projected separations of at least 20
\hkpc, i.e.  at distances where there is no overlap in the SDSS images
(typical half-light radii are on the order of 3 \hkpc) and
\textsc{GIM2D} does a good job of deblending the images (Simard et
al. in prep.).  However, the same is not true of the public SDSS
photometry, as we demonstrate below.

\begin{figure}
\centerline{\rotatebox{270}{\resizebox{6cm}{!}
{\includegraphics{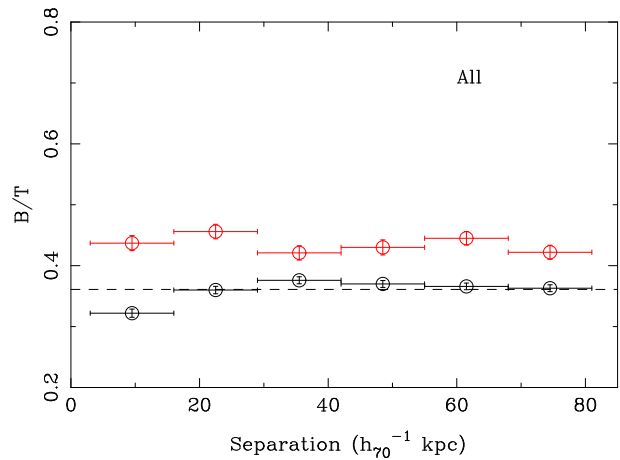}}}}
\caption{\label{bt_fig} Bulge fractions ($r$-band) for pair galaxies
(red points) with $\Delta v < 200$ \kms\ in all environments.  
Black points are control galaxies where the separation indicates
the $r_p$ of the pair to which it is matched.   The horizontal dashed
line shows the median B/T values for control galaxies. }
\end{figure}

\begin{figure*}
\centerline{\rotatebox{270}{\resizebox{13cm}{!}
{\includegraphics{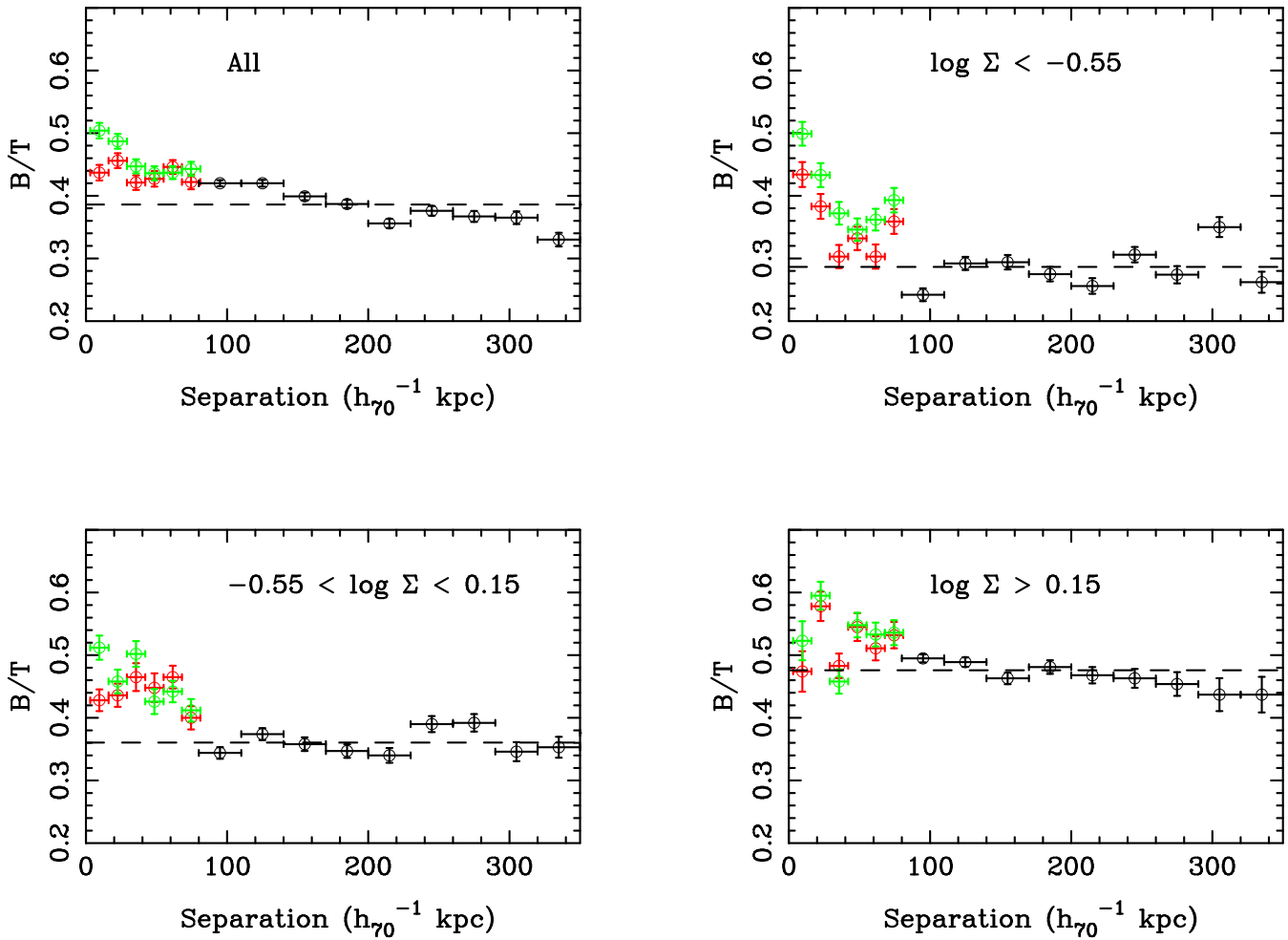}}}}
\caption{\label{bt_fig_wide} Bulge fractions ($r$-band) for pair
galaxies (red points) with $\Delta v < 200$ \kms\ as a function of
projected separation.  Black points are control galaxies where the
separation is the $r_p$ to the nearest neighbour in the DR7.  Green
points also show pair galaxy B/T ratios, but derived from the original
SDSS photometry where issues with the background and segmentation
images can lead to fitting and photometry errors in crowded environments.
The horizontal dashed line shows the median B/T values for control
galaxies.  The upper left panel shows the full sample, other panels
are split by local density $\log \Sigma < -0.55$ (upper right), $-0.55
< \log \Sigma < 0.15$ (lower left) and $\log \Sigma > 0.15$ (lower
right).  }
\end{figure*}

Before investigating the dependence of B/T in different environments,
in Figure \ref{bt_fig} we show the bulge-to-total (B/T) fractions in
the $r$-band for galaxies for all values of $\Sigma$.  Figure
\ref{bt_fig} ostensibly confirms previous reports of higher bulge
fractions at projected separations $r_p < 30$ \hkpc\ (similar trends
are seen in the $u$ and $g$-band).  However, by plotting the B/T as a
function of of $r_p$, we see the puzzling result that the bulge
fraction does not join the control galaxies at wider separations.
Recall that for every pair galaxy there are four control galaxies
matched in stellar mass, redshift and $\Sigma$.  If there is a
`typical' B/T at a given $\Sigma$, we would expect the pair values to
tend to this value at wide separations and for this value to be
reflected in the control galaxies.  However, Figure \ref{bt_fig} shows
that the pair galaxies have systematically higher B/T than the control
even at 80 \hkpc. 

To understand the origin of the offset in B/T between pairs and
control we identify the nearest galactic neighbour in the
spectroscopic catalogue (DR7) for the sample of control galaxies (no
companion within 80 \hkpc\ and 200 \kms).  The DR7 is used for this
test since we are not restricted by the requirement of
cross-correlating with the cluster catalogue (only available for DR4).
This allows us to extend the B/T versus separation plot out to larger
values of $r_p$ \textit{using the control galaxies}, see Figure
\ref{bt_fig_wide}. Note that in Figure \ref{bt_fig_wide} the
separation of the control galaxies is the true projected distance to
its nearest neighbour, whereas in Figure \ref{bt_fig} the separations
of the control galaxies refer to the $r_p$ of the pair galaxy to which
it is matched.  Figure \ref{bt_fig_wide} shows that (for the full
sample, top left panel) there is a monotonic dependence of B/T on near
neighbour distance over hundreds of kpc, an effect that is driven
largely by galaxies in high density environments (lower right panel).
The B/T of galaxy pairs in high densities is consistent with a
continuation of the bulge fraction scaling seen out to 400 \hkpc.  The
trends in Figure \ref{bt_fig_wide} are probably manifestations of the
well-known morphology-density relation (Dressler 1980) where the
distance to the nearest neighbour is a coarse indicator of local
environment.  The control galaxies are therefore offset in Figure
\ref{bt_fig} because B/T has not reached a fiducial value at $r_p <$
80 \hkpc, but continues to decline, due to the dependence of B/T on
both $r_p$ and $\Sigma$.  We note that this effect may contribute to
the finding that pairs have a higher early type fraction (Deng et al
2008), in addition to a bias towards higher density environments
(Barton et al. 2007).  A simple comparison of B/T in all pair galaxies
(out to 80 \hkpc) compared to the control would also indicate
misleadingly high bulge-fractions, which were not associated with
triggered star formation.

\begin{figure*}
\centerline{\rotatebox{270}{\resizebox{13cm}{!}
{\includegraphics{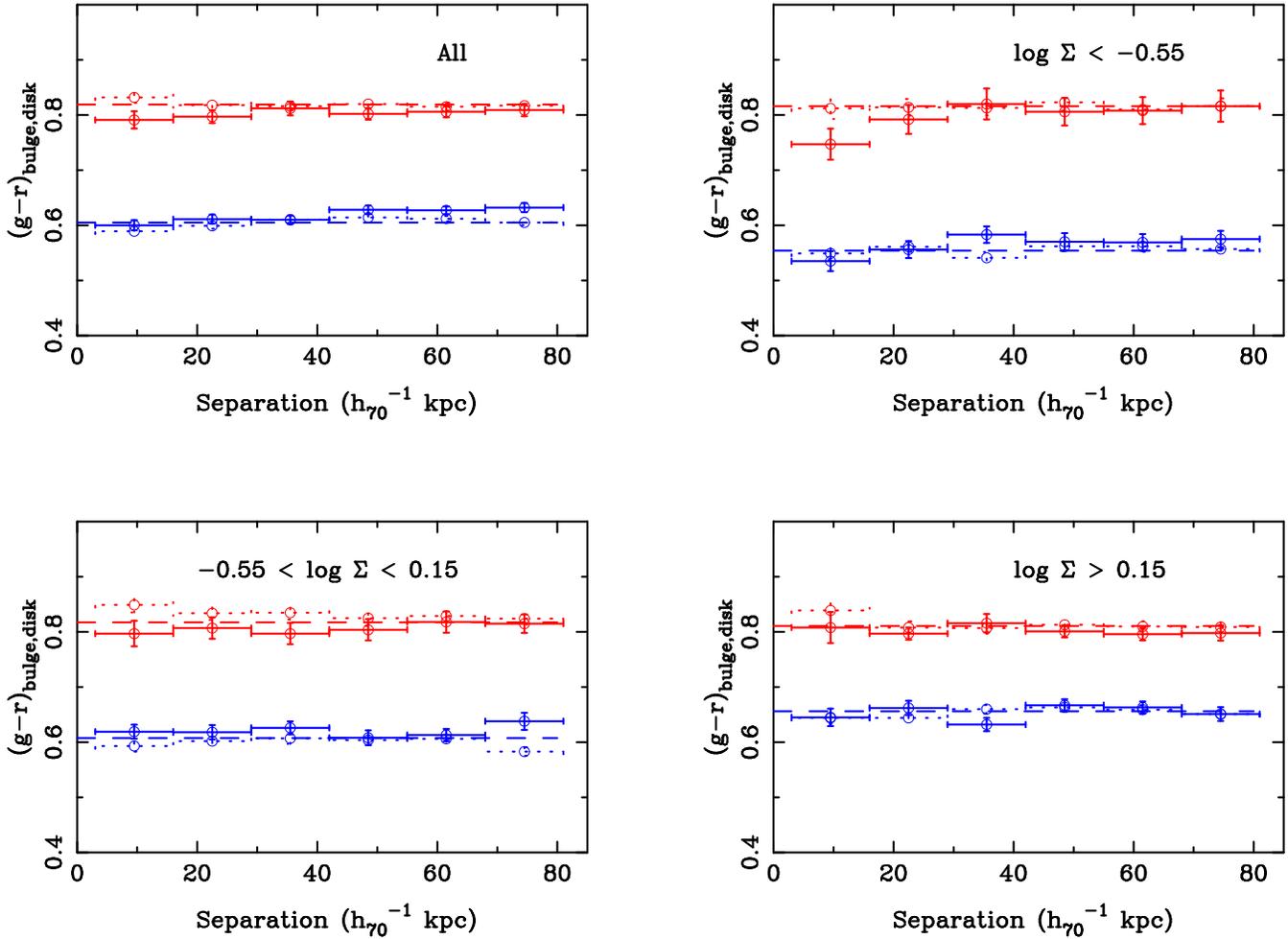}}}}
\caption{\label{gmr_sep_bulge} Median $g-r$ bulge (red) and disk (blue)
colour for pair galaxies
with $\Delta v < 200$ \kms\ as a function of projected separation. The
upper left panel shows the full sample, other panels are split by
local density $\log \Sigma < -0.55$ (upper right), $-0.55 < \log \Sigma
< 0.15$ (lower left) and $\log \Sigma > 0.15$ (lower right).  The
median value of $g-r$ colours for the appropriate density control
galaxies are shown as horizontal dashed lines and for individual bins in
dotted lines.}
\end{figure*}

Figure \ref{bt_fig_wide} shows that the typical B/T is higher in
higher density environments, as expected as the population becomes
more dominated by early-type galaxies.  In addition, there is a clear
trend, at all values of $\Sigma$, for the bulge fractions in galaxy
pairs to increase towards small separations. Taken at face value, this
result indicates that galaxy-galaxy interactions increase the bulge
fraction regardless of environment.  However, extending our analysis
to wider separations has demonstrated that local morphology-density
relations may contribute to higher bulge fractions independently of
ongoing interactions.  At high $\Sigma$, the enhanced B/T at $r_p <
80$ \hkpc\ is actually a smooth continuation of the trend at wider
separations and may therefore not be connected to the current
interaction.  This interpretation is consistent with our finding
(Figure \ref{sfr_sep}) that there is little triggered star formation
in high density environments.  Conversely, in the low density
environments there is a striking enhancement in B/T at separations
$r_p < 30$ \hkpc, a trend that is also seen slightly more modestly
(but at wider separations) at intermediate values of $\Sigma$.  Again,
this agrees with the results of Figure \ref{sfr_sep} that shows that
triggered star formation occurs more prolifically at lower values of
$\Sigma$.  Our results therefore support a picture of centrally
concentrated star formation that is effective primarily in
low-to-intermediate density environments.

We also demonstrate in Figure \ref{bt_fig_wide} the effect of crowding
on the SDSS photometry.  Bulge fractions derived from the original
SDSS images are shown in green and fits to the improved background and
segmentation images are shown in red.  It can be seen that the B/T
can be over-estimated at small separations when the original SDSS
images and segmentation maps are used.  

\subsection{Bulge and disk colours}

\begin{figure*}
\centerline{\rotatebox{270}{\resizebox{13cm}{!}
{\includegraphics{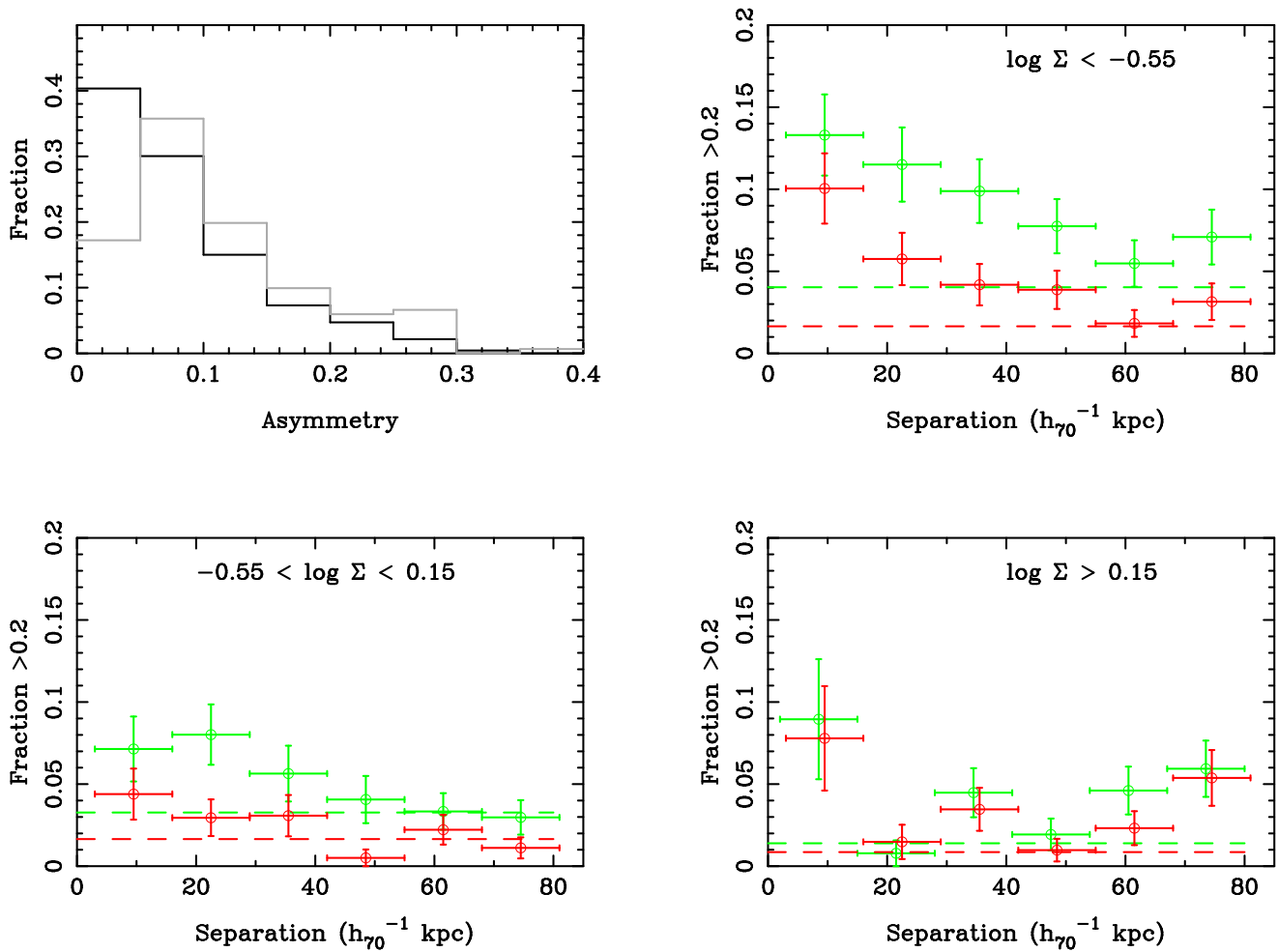}}}}
\caption{\label{asym_fig} Top left panel: distribution of $g$-band
asymmetries for galaxy pairs with $\Delta v <$ 200 \kms\ and in low
density environments ($\log \Sigma < -0.55$).  The grey histogram is
for galaxies in the innermost $r_p$ bin ($3<r_p<16$ \hkpc) and the
black histogram is for the wide separation bin with $69<r_p<80$ \hkpc.
The other panels show the fraction of galaxies with a close companion
within $\Delta v < 200$ \kms\ whose asymmetry $>$0.2 as a function of
projected separation.  Green and red points indicate asymmetry for the
$g$ and $r$ filters respectively.  Panels are split by local density:
$\log \Sigma < -0.55$ (upper right), $-0.55 < \log \Sigma < 0.15$
(lower left) and $\log \Sigma > 0.15$ (lower right).  The fraction of
control galaxies with asymmetry $>$ 0.2 (by $\Sigma$ and filter) are
shown as horizontal dashed lines. In the lower right panel, the green
points have been offset horizontally by a small amount for clarity.}
\end{figure*}

If star formation is located primarily in the galaxy centres, colour
gradients might be expected to arise from a dominant blue population
in the inner regions.  Indeed, Barton et al. (2003)
found that in their sample of 190 galaxies in pairs and compact
groups many showed bluer colours in their centres.  Kewley
et al. (2006) measure a colour offset of the inner versus outer disk
and find that this correlates with a decrease in metallicity, supporting
the picture of gas inflow as a trigger.

Since \textsc{GIM2D} decomposes the galaxy into a bulge and disk
component and calculates the relative flux in each, it is possible to
examine not only the fractional distribution of the light in the two
components, but also the colours of the disk and the bulge.  Although
$u-r$ provides one of the stronger diagnostics of star formation,
there are several limitations of the $u$-band in the SDSS make this a
less practical option.  For example Baldry et al. (2005) discuss the
fact that the survey limit approaches the $u$-band magnitude for many
`typical' galaxies and that several background issues can affect the
Petrosian magnitudes.  A related issue is that a non-negligibile
fraction of our sample is not detected in the $u$-band.  In Figure
\ref{gmr_sep_bulge} we show the bulge and disk $g-r$ colours for pairs
with $\Delta v < 200$ \kms\ as a function of projected separation for
different cuts in $\Sigma$.  The error in the bulge and disk colours
is propogated from errors in the bulge and disk magnitudes and the
error in the bulge fraction.  Only galaxies with errors less than 0.5
mags in the bulge or disk magnitudes were considered.  Figure
\ref{gmr_sep_bulge} shows that the pairs sample as a whole exhibits a
small bluing of the bulge, but not the disk, at small separations.
When considered as a function of $\Sigma$, it is the bulges in low
density environments that show a significant bluing at small
separations by $\sim$ 0.1 mag. There are no convincing colour changes
in the bulge at intermediate or high densities or in the disk at any
$\Sigma$.  This is further evidence that galaxy-galaxy interactions
are most effective at triggering central star formation in low density
environments.

\subsection{Asymmetry}

Star formation may be expected to increase the asymmetry of galaxies,
due to the clumpy nature of its distribution.  Several papers have
documented the increased asymmetry of galaxies in close pairs, e.g.
Patton et al. (2005) and De Propris et al. (2007).  However, the
change in a galaxy's smoothness may depend on the competing effects
of star formation and dust. Lotz et al.  (2008) used simulations of
galaxies with and without dust to show that extinction around star
forming regions can actually smooth out the light distribution. Moreover,
asymmetry in interacting galaxies can be introduced even in the
absence of star formation due to the formation of tidal features.
A final complication is that asymmetries may only last for a fraction
of the full interaction, and the degree of asymmetry depends on
internal properties (such as gas fraction) as well as the geometry
of the merger (Lotz et al. 2008).  The fraction of galaxies which
exhibit asymmetry is therefore a conservative measure of interactions.

The top left panel of Figure \ref{asym_fig} shows the distribution of
asymmetries in the smallest separation pairs at low $\Sigma$, and a
comparison of more widely separated pairs.  This demonstrates that the
change in asymmetry at small $r_p$ is not simply a wholesale shift to
larger values.  The change is a two-fold loss of very smooth
morphologies and the development of a more extended tail of
high values.  Therefore, we follow Patton et al. (2005) and use
fraction of galaxies with asymmetries greater than a threshold value,
rather than a central value estimator such as the median or mean.

The remaining three panels of Figure \ref{asym_fig} show the fraction of
galaxies with asymmetry greater than 0.2 in the $g$ and $r$ bands as a
function of projected separation.  For clarity, we do not plot the
control galaxies as a function of separation, but simply their high
asymmetry fraction calculated for the appropriate filter and $\Sigma$.
The asymmetric fraction increases towards small separations; this is true for
all environments, even though there is little evidence for strong star
formation in the high $\Sigma$ sample.  This indicates that the
increased asymmetry is not (solely) due to clumpy star formation,
since we see no other evidence for star formation at high $\Sigma$,
e.g. from SSFRs or bulge colours.  The asymmetries in high density
environments are more likely to be associated with tidal features.
Further clues to the source of the asymmetry comes from looking at the
contrast between different filters.  At low densities, the increase in
asymmetry at small $r_p$ is more enhanced in the $g$-band than
the $r$-band, consistent with expectations from star formation.  At
high densities, the $g$ and $r$-band show equal asymmetry
enhancements, indicating that the colour of the residuals is not
changing.  This is expected if the asymmetry is caused by a
re-distribution of the existing stellar population, e.g. through tidal
disruption.  Taken together, these results indicate that galaxy-galaxy
interactions occur at all densities, but that triggered star formation
is restricted to the lower density environments.  One explanation for
this result is the tendency for galaxies in high density environments
to be relatively gas poor, and therefore lack the gas content to feed
new star formation.  We return to this suggestion in the context of
other recent results in Section 5.

\section{The Environment of Galaxy Pairs: Cluster Membership.}\label{member_sec}

In this section we investigate what larger structures galaxy pairs are
embedded in and how this might affect interactions.  Specifically, we
consider more extreme structural membership, by investigating galaxy
pairs located in a previously compiled cluster catalogue.  The cluster
catalogue is based on the compilation of von der Linden et al. (2007)
who refined the C4 algorithm (Miller et al. 2005) and applied it to
the DR4.  The original C4 catalogue identifies galaxy overdensities in
seven-dimensional position and colour space.  The DR2 catalogue of
Miller et al. (2005) is estimated to be 90\% complete and 95\% pure.
In applying the C4 algorithm to the DR4, von der Linden et al.  (2007)
made a number of improvements, such as removing clusters with unusual
velocity distributions, or the calculation of radius and velocity
dispersion does not converge.  Clusters without a clearly identifiable
brightest cluster galaxy were also removed.  The final von der Linden
catalogue contains 625 clusters containing a total of 18,100 galaxies.
After applying the additional criteria described in Section
\ref{sample_sec} and requiring that the galaxy is associated with a
unique cluster, 14,857 galaxies remain in our cluster catalogue.

\begin{figure*}
\centerline{\rotatebox{270}{\resizebox{13cm}{!}
{\includegraphics{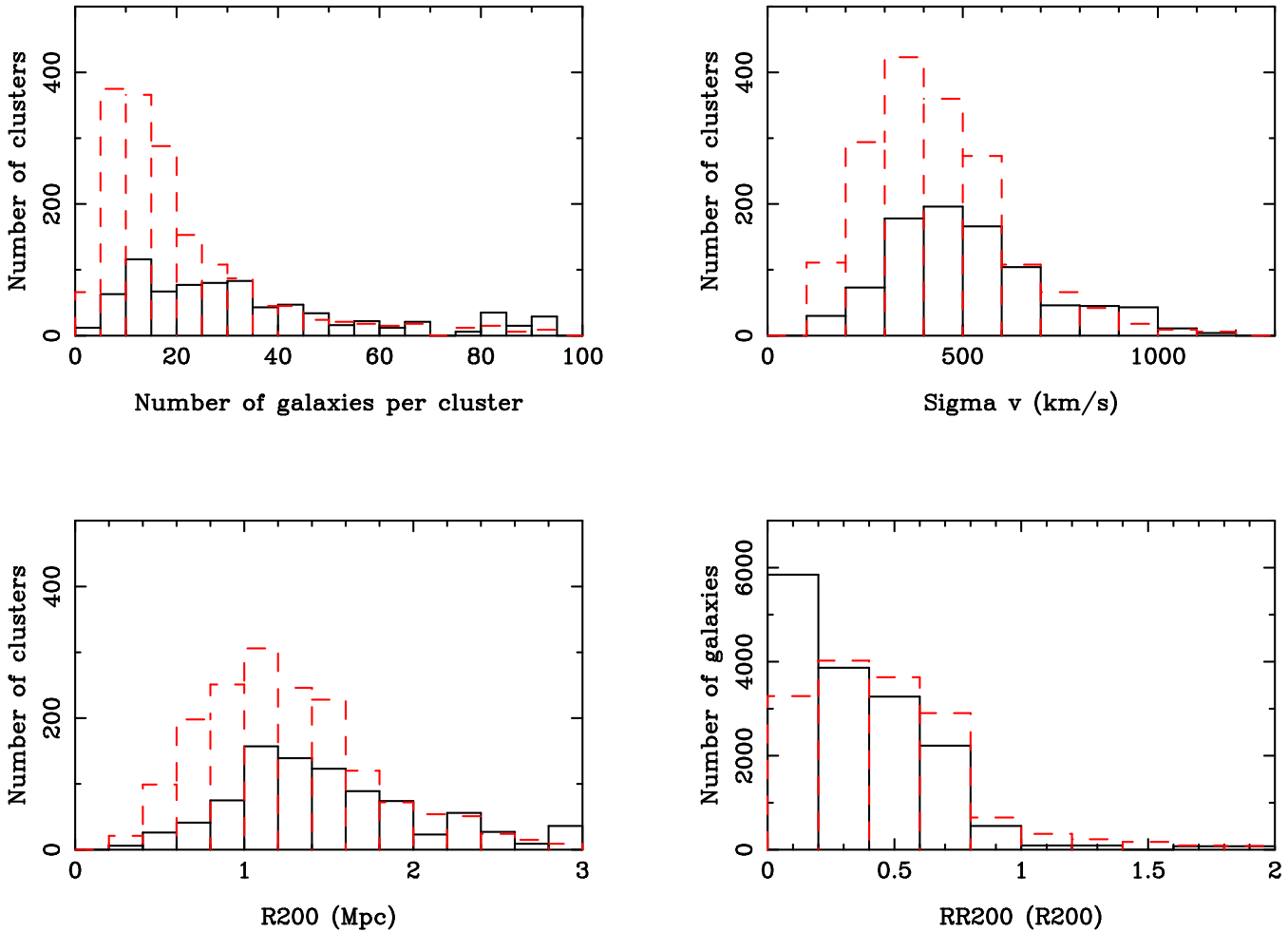}}}}
\caption{\label{cluster_dist} Histograms of cluster properties for
galaxies in full C4 sample (red dashed line) and the clusters in
which close pairs have been identified (solid black line). The lower
right panel shows the distributions of clustercentric radii for all C4
galaxy members (red dashed histogram) and those of close pairs (solid
black line).  The red dashed histograms of cluster properties have
been scaled up by a factor of 2 and the solid black line for the RR200
distribution of cluster pairs has been scaled up by a factor of
17 for presentation purposes.}
\end{figure*}

Of the 5784 galaxies in our pairs sample, 896 (15\%) have been
identified as a cluster galaxy in the C4 compilation.  In practice,
there are likely to be many pairs in clusters that do not appear in
our sample due to fibre collision incompleteness.  For example, 58\%
of galaxies in the VIVA Virgo cluster sample have a close companion
within 80 kpc and 300 \kms\ (Chung, 2009, private communication).
However, the majority of galaxies with a close companion in our
SDSS-selected sample are not identified as cluster members.  In Figure
\ref{cluster_dist} we investigate where in the cluster the 896 pairs
are located and whether or not they inhabit a particular subset of C4
clusters.  The Figure shows the distribution of cluster size (R200),
velocity dispersion ($\sigma_v$) and number of galaxy members for all
clusters in our catalogue, and those clusters in which a close pair
has been identified.  Also shown is the distribution of clustercentric
radii (RR200) in units of R200 for all C4 catalogue members and the
896 members with a close companion.  Figure \ref{cluster_dist} shows
that the clusters in which we have identified close pairs have larger
mean velocity dispersions, smaller clustercentric distances, R200
sizes and number of galaxy members than the full C4 sample.

\begin{figure}
\centerline{\rotatebox{0}{\resizebox{8cm}{!}
{\includegraphics{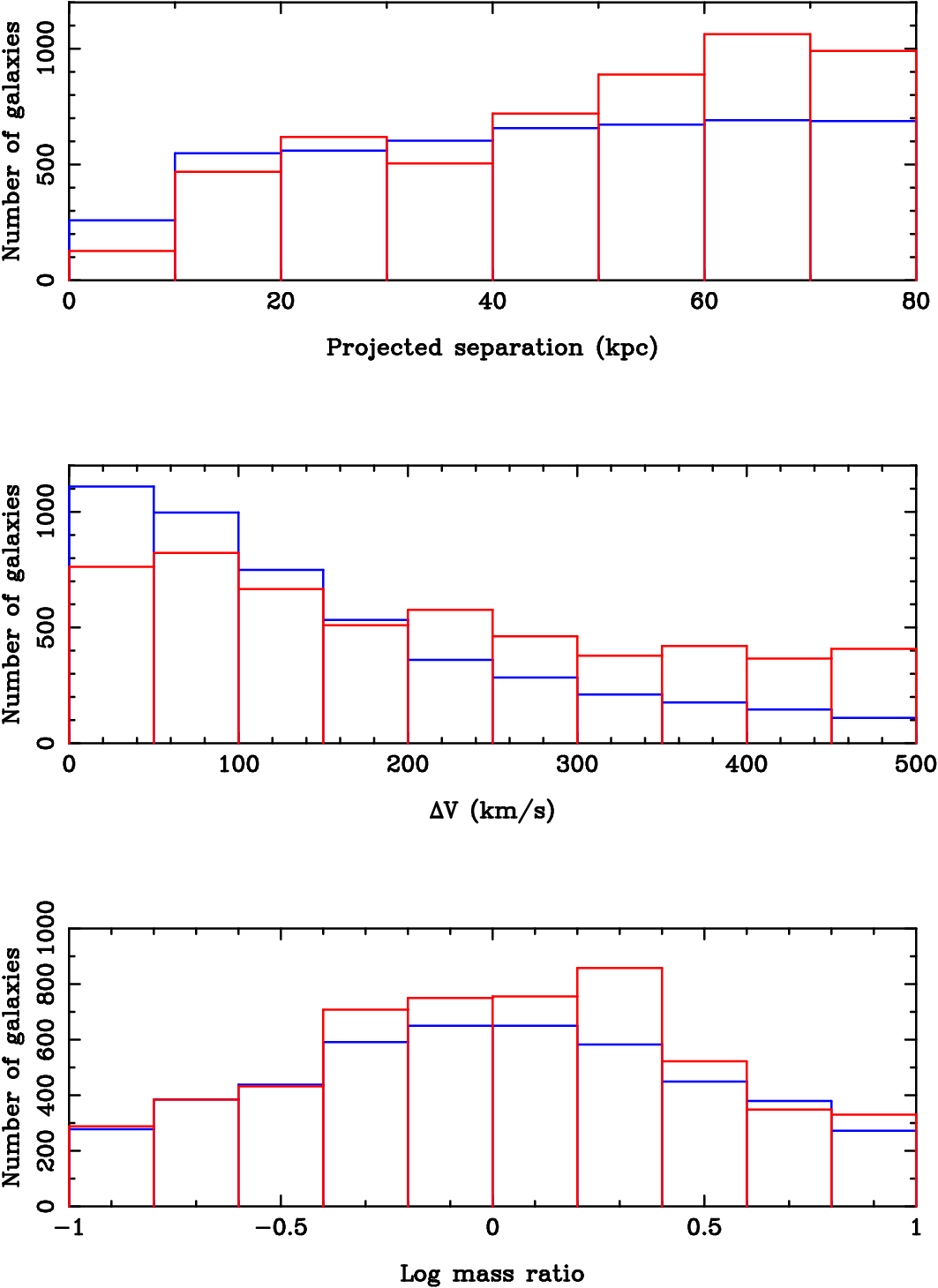}}}}
\caption{\label{pairs_mem_histo} Histograms of pairwise properties for
galaxies in clusters (red) and the field (i.e., non-clusters, blue). 
The cluster histogram has been scaled by a factor of 5 for presentation
purposes.}
\end{figure}

\begin{figure}
\centerline{\rotatebox{270}{\resizebox{6cm}{!}
{\includegraphics{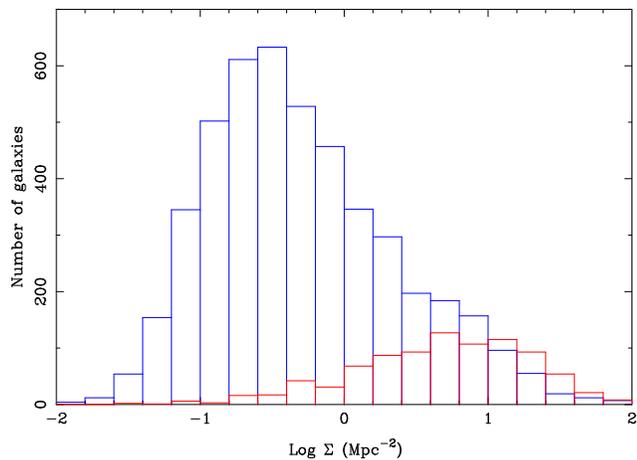}}}}
\caption{\label{cluster_sigma} Distributions of projected density for
pair galaxies in clusters (red),  and the field (i.e., non-clusters, blue).}
\end{figure}

In Figure \ref{pairs_mem_histo} we show the pairwise properties of our
galaxy pairs sample colour coded according to whether or not they appear
in the cluster catalogue.  We see a similar trend for membership 
classification as for
environment parameterization by $\Sigma$ (Figure
\ref{pairs_sig_histo}), i.e. a tendency towards wider projected
separations and large $\Delta v$ in the clusters than the 
`field' pairs.   The connection
between high $\Sigma$ and cluster membership is demonstrated
explicitly in Figure \ref{cluster_sigma}, where the projected
densities of field and cluster pairs are plotted.  82\% of pairs
in clusters are in the highest density tertile ($\log \Sigma > 0.15$).
However, 62\% of the highest density tertile pairs are non-cluster.

\subsection{Mergers in clusters}

Despite the locally high densities, clusters are often considered
inhospitable for galaxy mergers due to the large velocity dispersions
involved.  Instead, the group environment is assumed to be the prime
location for mergers. For example, Wilman et al. (2009) find S0s
dominate in group environments, and Just et al. (2010) find the S0
fraction evolves most strongly in intermediate velocity dispersion
structures.  These results imply that groups are an important location
for spiral mergers, in agreement with the idea of group
`pre-processing' (e.g. McGee et al. 2009).  However, there is evidence
that interactions (if not mergers) do nonetheless occur in clusters.
For example, Chung et al. (2007) found that 6/7 of the late-type Virgo
cluster galaxies in their sample which exhibit long HI gas tails also
had a close companion within 100 kpc and 100 \kms.  It was suggested
that tidal interactions may contribute to the efficiency of ram
pressure stripping.
 
We investigate whether interactions are elevated in clusters by
comparing the asymmetry of galaxies in close pairs ($\Delta v < 300$
\kms\ and $r_p < 20$ \hkpc) in the C4 catalogue with wide pairs
($\Delta v > 400$ \kms\ or $r_p > 40$ \hkpc) in clusters.  Since we do
not expect the wide pairs to be effective in galaxy-galaxy
interactions, it is possible to use them as a comparison sample.  Lin
et al. (2010) have shown that although the galaxy pair fraction
increases with density, once the higher fraction of interlopers is
accounted for, the fraction of pairs that will merge actually
decreases with density.  

\begin{figure}
\centerline{\rotatebox{270}{\resizebox{6cm}{!}
{\includegraphics{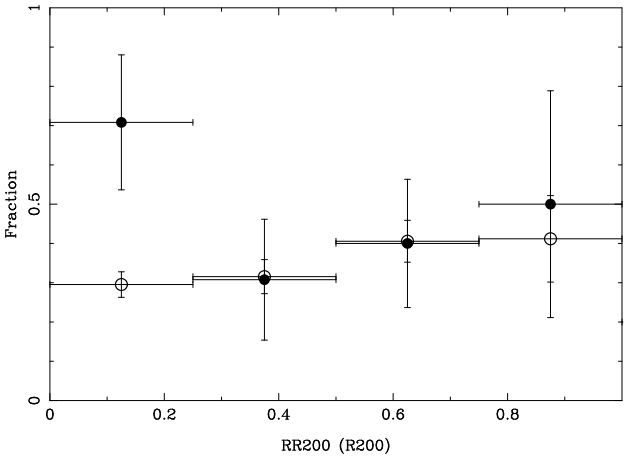}}}}
\caption{\label{asym_RR200_fig} Fraction of galaxies with asymmetry $>$ 0.05
in the $r$-band for close
pair galaxies ($\Delta v < 300$ \kms\ and $r_p < 20$ \hkpc, solid
points) and wide pairs ($\Delta v > 400$ \kms\ or $r_p > 40$ \hkpc,
open points) and as a function of clustercentric distance.  Only
galaxy pairs that are matched in the C4 cluster catalogue are
included. }
\end{figure}

In Figure \ref{asym_RR200_fig} we show the
fraction of galaxies with $r$-band asymmetries $>$ 0.05 as a function
of clustercentric distance for the close and wide pairs.  There is a
mild trend towards smoother morphologies at smaller clustercentric
distances in the wide pairs, as would be expected from a higher
elliptical fraction in the higher local densities of the cluster core.
Conversely, there is a sharp increase in the fraction of asymmetric
close pairs within 0.25 R200.  A 2D KS test on asymmetry and RR200
rules out the null hypothesis that the close and wide pair samples
were drawn from the same distributions at 99.99\% confidence.  The
close pairs exhibit no trend of B/T with RR200, so the high asymmetric
fraction is not due to a higher spiral fraction in the inner RR200
bin.  There is also no trend of SFR with RR200, indicating that the
higher asymmetric fraction is associated with tidal disruption rather
than star formation.  The local density is a strong function of
clustercentric distance, with median values ranging from log $\Sigma
\sim 1$ at 0.1 RR200 to log $\Sigma = -0.25$ at 0.9 RR200 for the
close pairs.  The high asymmetric fraction of close pairs in the inner
regions of clusters shown in Figure \ref{asym_RR200_fig} is therefore
consistent with the strong increase in asymmetric fraction found at
$r_p < 20$ \hkpc\ and log $\Sigma > 0.15$ in Figure \ref{asym_fig}.

\section{Summary \& Discussion}\label{summary_sec}

By using a sample of galaxies with a companion ($r_p < 80$ \hkpc,
$\Delta v < 500$ \kms) selected from the SDSS DR4, combined with
metrics of colour, morphology and star formation, we have examined in
what environments mergers occur and what the result of the interaction
is.  Below we summarise our principle findings.

\begin{enumerate}

\item The pairwise properties of galaxies with a close companion
depend on environment.  Low density environments have fractionally
more pairs with low projected separations and low relative velocities
than high density environments.  Conversely, high density enviornoments
selected either by 4+5th nearest neighbour projected density, or cluster
membership  are characterised by wider separations and larger values of
$\Delta v$.  The imposed velocity cut for a given 
pairs sample therefore also acts as a local density selection.

\item The range of stellar mass ratios does not depend on environment
\textit{for complete samples}.  We select a subset of pairs with
$\Delta v < 200$ \kms\ to study the effects of galaxy-galaxy
interactions.

\item Galaxies in all environments, as defined by $\Sigma$, may
undergo interactions if they have a close companion, as shown by
increased asymmetry in the $g$ and $r$-band.  However, in clusters, we
find evidence for galaxy-galaxy interactions (higher asymmetry) only
in the cluster centre.  

\item Triggered star formation occurs mainly in low density
environments where gas fractions are expected to be typically
higher. The star formation rate is also seen to increase, but to a lesser
extent, in intermediate density environments.  Bluer bulge $g-r$
colours indicate that the star formation occurs mainly in the central
regions of galaxies.

\item In high density environments, mergers still occur, but
they are mainly without star formation.  This is demonstrated by the
consistent colour change in the galaxy asymmetries in the $g$ and $r$
filters.  This is also true for pairs
in clusters where we see higher asymmetries without an increase in star
formation rate.  

\item  In high density environments, the bulge fraction is a monotonically
increasing function of nearest neighbour separation out to at least 400
\hkpc.  At lower galaxy densities, there is a clear
enhancement of B/T at separations $<$ 30 \hkpc\ which we interpret
as the signature of the central star formation proposed above.

\end{enumerate}

These results bring together three themes that have been emerging
separately in the literature.  First, the now well-established
enhancement in star formation rate in galaxy pairs, which apparently
dominates in lower density environments.  Indeed, Lin et al. (2010)
have shown that most `wet' mergers (i.e. those between two gas-rich
galaxies on the blue cloud) occur at low densities.  Our data clearly
show that this star formation occurs preferentially in the central
part of the galaxy, rather than in its extended disk.  Secondly, the
emerging result that massive early-type galaxies are also hotbeds of
merger activity.  For example, Tal et al (2009) imaged a complete
sample of nearby luminous ellipticals and found evidence of
asymmetries in 73\% of them.  Despite the apparent prevalence of
recent interactions, these massive galaxies showed no signs of
enhanced star formation.  A similar conclusion was drawn by McIntosh
et al. (2008) who find no difference in colours or concentrations
amongst massive major mergers in groups (although as we note above,
the overall colour distribution may be relatively insensitive to
modest enhancements in star formation) and Tran et al. (2008) who find
only old stellar populations in their disturbed group galaxies.  High
density environments become systematically more dominated by masssive
red galaxies (Kauffmann et al. 2004; Balogh et al. 2004; Baldry et
al. 2006) and galaxy-galaxy interactions therein become more dominated
by `dry' mergers (Lin et al. 2010).  This is supported by our finding
that at large $\Sigma$ there are enhanced colour dependent asymmetries
in galaxies with small projected separations, but with no accompanying
change in SFR or bulge colour.  Finally, we have addressed where
mergers take place.  In general, mergers are considered to be rare in
relaxed clusters (e.g. Makino \& Hut 1997).  Observationally, massive
cluster galaxies show a lower incidence of interaction (through
asymmetries) than galaxies in groups or the field (McIntosh et
al. 2008; Liu et al 2009; Tal et al 2009).  In a study of the A901/902
supercluster, Heiderman et al. (2009) identified 13 morphologically
disturbed galaxies with enhanced star formation, all outside the
cluster core.  Our study has selected close pairs of galaxies as
sites of possible mergers and is complimentary to studies which
select interacting galaxies based on visible asymmetries (e.g.
Lotz et al. 2010).  We find evidence for tidal interactions even in the
cluster core, although with no signs of associated star formation.

In summary, galaxy-galaxy interactions appear to be ubiquitous in the
local universe.  However, despite
their ubiquity, the observational manifestation of the interaction,
such as triggered star formation, colour changes and tidal disruption,
depends markedly on environment. 

\section*{Acknowledgments} 

We are grateful to Anja von der Linden and the MPA/JHU group for
access to their data products and catalogues (maintained by Jarle
Brinchmann at http://www.mpa-garching.mpg.de/SDSS/).  Aeree Chung
generously shared results of the VIVA survey in advance of
publication.  Thanks to Josefa Perez for comments on a previous draft.
SLE and DRP acknowledge the receipt of NSERC Discovery grants which
funded this research.

\end{document}